\documentclass{appolb}
\usepackage{graphicx}
\usepackage{amsmath}
\usepackage{cite}
\usepackage{color}
\graphicspath{{figures/}}
\usepackage[utf8]{inputenc}
\usepackage[english]{babel}

\title{Determining crucial factors for the popularity of scientific articles}
\author{Robert Jankowski \footnote{e-mail: robert.jankowski3.stud@pw.edu.pl} and Julian Sienkiewicz \footnote{corresponding autor, e-mail: julian.sienkiewicz@pw.edu.pl}
\address{Faculty of Physics, Warsaw University of Technology,\\ Koszykowa 75, PL-00662 Warszawa, Poland}}

\begin{document}

\maketitle

\maketitle
\begin{abstract}
Using a set of over 70.000 records from PLOS One journal consisting of 37 lexical, sentiment and bibliographic variables we perform analysis backed with machine learning methods to predict the class of popularity of scientific papers defined by the number of times they have been viewed. Our study shows correlations among the features and recovers a threshold for the number of views that results in the best prediction results in terms of Matthew's correlation coefficient. Moreover, by creating a variable importance plot for random forest classifier, we are able to reduce the number of features while keeping similar predictability and determine crucial factors responsible for the popularity. 
\end{abstract}

\section{\textsc{INTRODUCTION}}
The single most popular bibliometric criterion for judging the impact of scientific papers is the number of citations received, commonly known as ``citation count''. However, due to a rather indirect discipline dependence \cite{didegah2013}, this metric alone can be in many cases unreliable.  Many services have created their own metrics to determine the most popular article which shows a complex landscape of measures \cite{bollen2009}. Google Scholar\footnote{https://scholar.google.com/} puts weight especially on citation count and words included in a document's title. As a consequence, the first search results are often highly cited articles. On the other hand, Public Library of Science\footnote{https://www.plos.org/} takes into account the number of HTML page views and PDF downloads. In Scopus\footnote{https://www.scopus.com/}, there are a few different article-level metrics: (i) Scholary Activity (Mendeley readers), (ii) Scholary Commentatory (Blogs, Wikipedia), and (iii) Mass Media (Twitter, Facebook, Reddit), which brings this service closer to so-called ``altmetrics'' \cite{Glaenzel2015} which has lately become one of the key components used in evaluating scientific papers.

There is an overwhelming motion among scientists \cite{paiva2012,letchford2015,rostami2014} and in particular in the popular media overages to connect the popularity of articles to some of their textual features in a simple way. A running example is to directly state that shorter titles increase the chance of a scientific article to be cited \cite{deng2015}. A counterexample to this simplified approach is a recent paper \cite{Sienkiewicz2016}, where authors performed an investigation of how different textual properties of scientific papers affect the number of citations they acquire. Using a set of over 4.3 million and applying quantile regression in order to examine different regimes of citations it has been shown that in most journals, short titles correlate negatively with citations only for the most cited papers. On the other hand, for typical papers, this relation is usually absent. It has also been noticed that depending on the journal the results may vary, which emphasizes the nonlinearity of the relationship.

In this work, we approached the problem from a different point of view. The main distinction is that we not only intend to explore the popularity aspects but also build models which would, in turn, allow predicting whether the paper would be popular based on available features. In addition to that, as considering just a single feature (e.g., the length of the title) at a time might not be sufficient to cover complex relations among different features, we decided to use the machine learning (ML) approach. Moreover, this study takes into account a single journal (i.e., PLOS One), as compared to the previous investigation\cite{Sienkiewicz2016} in which authors used over 1500 different journals. Last but not least, we recorded the information about the number of views per month, in contrast to the sum of citation the article acquired like in the previous paper. This distinction seems to be important as almost everyone has a chance to view the page with the given article, but only people who are experts in a specific topic can cite it. Thus, the number of views is more a democratic measure.

The rest of the paper is organized as follows: first we describe in detail the data and show some exploratory data analysis to examine correlations among the features and distribution of views as well as its consequences. Then we introduce machine learning methods and two basic metrics that allow for quantification of predictability of tested methods. The fourth chapter contains the results of this study: the quality of predictions on the full dataset using selected classifiers and similar outcomes for reduced data. We end the paper with some concluding remarks.

\section{\textsc{DATA \& EXPLORATORY ANALYSIS}}\label{sec:sec2}
\indent
In this study we use two datasets: the first one has been downloaded using Public Library of Science automatic services (PLOS API)\footnote{http://api.plos.org} and it included the following information about each paper published in the PLOS ONE journal: (i) title, (ii) list of authors, (iii) abstract contents, (iv) full text contents. The downloaded data comprises around 140.000 records of data.
The second set was acquired using Article-Level Metrics data (PLOS ALM)\footnote{http://alm.plos.org} -- a service that keeps monthly statistics of views (i.e., visiting web-page with the given paper). After data cleaning, the dataset contained over 70.000 papers from 2003 to 2014.

\begin{table}[]
    \centering
\begin{tabular}{c|p{5cm}|p{6cm}}
\hline\hline
& property & comments\\
\hline
1 & number of words in the abstract, title and the full text (3) & -\\
\hline
2 & number of characters in the abstract, title and the full text (3) & -\\
\hline
3 & number of sentences in the abstract and the full text (2) & -\\
\hline
4 & number of authors (1) & - \\
\hline
5 & number of citation in the full text (1) & To calculate the number of citation the regular expression was used: $[number]$\\
\hline
6 & number of citations in the introduction and discussion (2) & See above comment\\
\hline
7 & valence and arousal for the title, abstract and full text (6) & Using dictionary provided by Warriner and el. \cite{valence}\\
\hline
8 & valence and arousal related to specific citation in the full text (2) & The sentiment was calculated taking into account 100 characters before and after the citation. \\
\hline
9 & valence and arousal related to specific citation after splitting full text into four parts (8) & The full text was divided into 4 parts; for each part positions of the citations were identified and then the sentiment was calculated as above. \\
\hline
10 & valence and arousal after spitting full text into four parts (8) & -\\
\hline
11 & number of views from date of a publication to 2014 (1) & -\\
\hline
\end{tabular}
    \caption{Features calculated for each article. Numbers in parentheses in the second column give the number of features coming from this property.}
    \label{tab:tab1}
\end{table}

Table \ref{tab:tab1} gathers all the features were calculated for item in the dataset (37 features in total). They are connected to different aspects of the scientific text: number of words, characters [and sentences] are purely \textit{length} properties.
Another widely utilized propriety of text is its \textit{emotional content} (sometimes called sentiment); here we use valence (i.e., the emotional sign of the text) and arousal which indicates the level of activation evoked by the text. These parameters have been used to quantify the collective behavior of online users \cite{Chmiel2011}, to model the evolution of online discussions in certain environments \cite{Sienkiewicz2013} or to predict the dynamics of Twitter users during Olympic Games in London \cite{Choloniewski2014}. In order to find the sentiment of the title and abstract the study by Warriner \textit{et al.} was used \cite{valence} where authors created a database containing nearly 14.000 English lemmas with valence and arousal values. Finally, the last property taken into account in this study are the \textit{references} appearing in the text. Here, we have been interested both in the plain number of citations, the part of the manuscript where they appeared (introduction, main part, discussion) as well as in their emotional context, i.e., the sentiment in the proximity of the given reference. This kind of analysis can help to judge the role of negative citations \cite{Catalini2015}.

\begin{figure}[h!]
      \centering
      \includegraphics[width=1.07\textwidth]{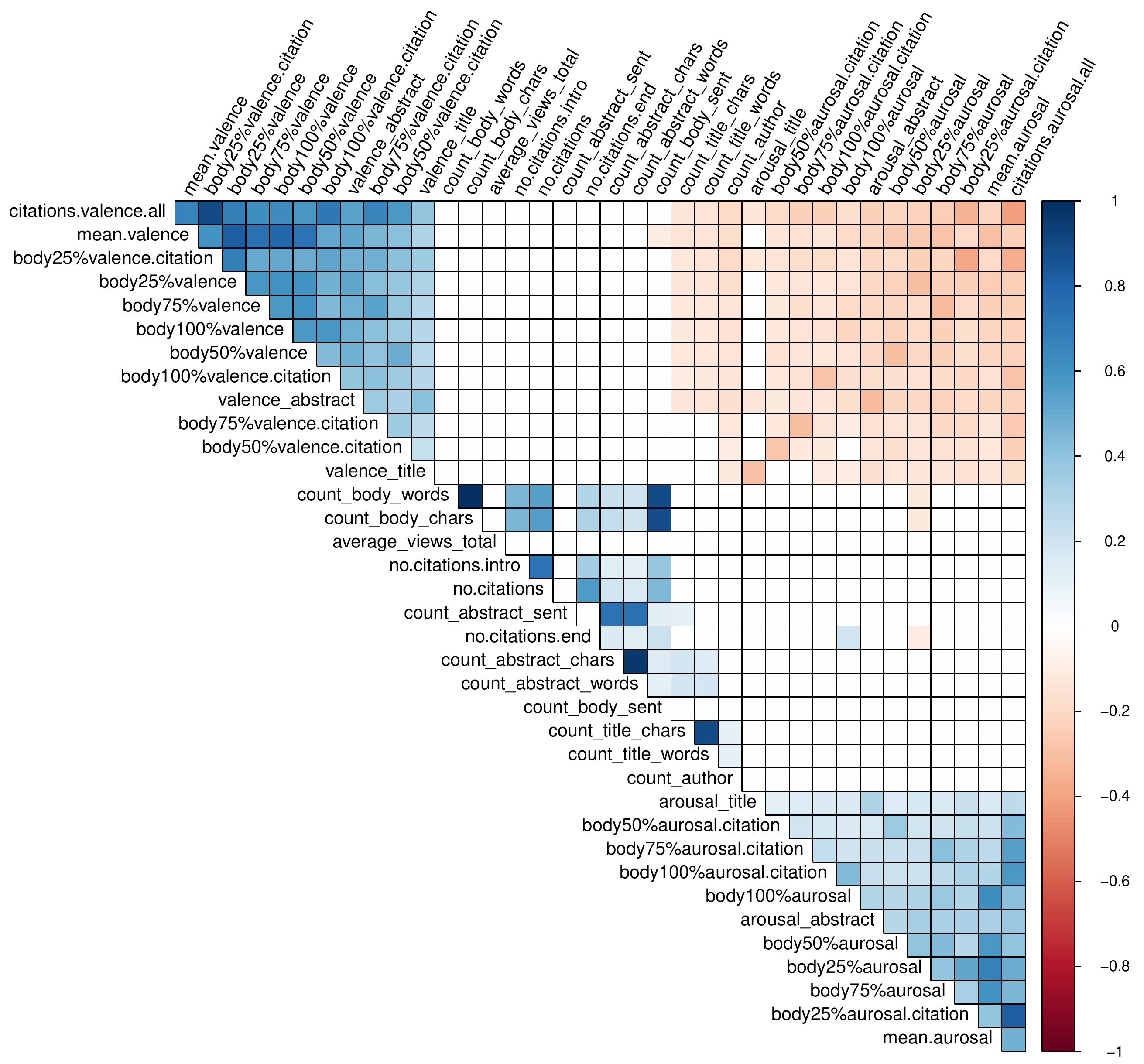}
      \caption{Correlation matrix of all features. To obtain better visual effect small correlation values ($r \in $$[-0.1, 0.1]$) were to zero.}
      \label{fig:correlation}
\end{figure}

Figure \ref{fig:correlation} shows a correlation matrix based all the features recognized in text: each cell of the matrix is simply calculated as a Pearson correlation coefficient of two features $X_i$ and $X_j$
\begin{equation}
\rho_{X_i,X_j} = \frac{\mathrm{cov}(X_i,X_j)}{\sigma_{X_i}\sigma_{X_j}}    
\end{equation}
where $\mathrm{cov}(X_i,X_j)$ is the covariance of features $X_i$ and $X_j$ and $\sigma_{X_i}$, $\sigma_{X_j}$ are, respectively, their variances. This shows an obvious, strong correlation between the number of words, characters, and sentences in the abstract, title and full text (middle part of the plot). Moreover, there is a noticeable relationship linking the number of citations and the number of words/characters in the full text. The valence features are correlated positively with each other (bottom-right part of the plot) as well as the arousal features (upper-left part of the plot) however, the correlation between the valence and arousal features is slightly negative (upper-right part of the plot). Interestingly, the features connected with sentiment are weakly or not correlated with \textit{length} properties. It is also notable that none of these features has any connections to the main observable of this study, i.e., the average number of views.

\begin{figure}[h!]
      \centering
      \includegraphics[width=.85\linewidth]{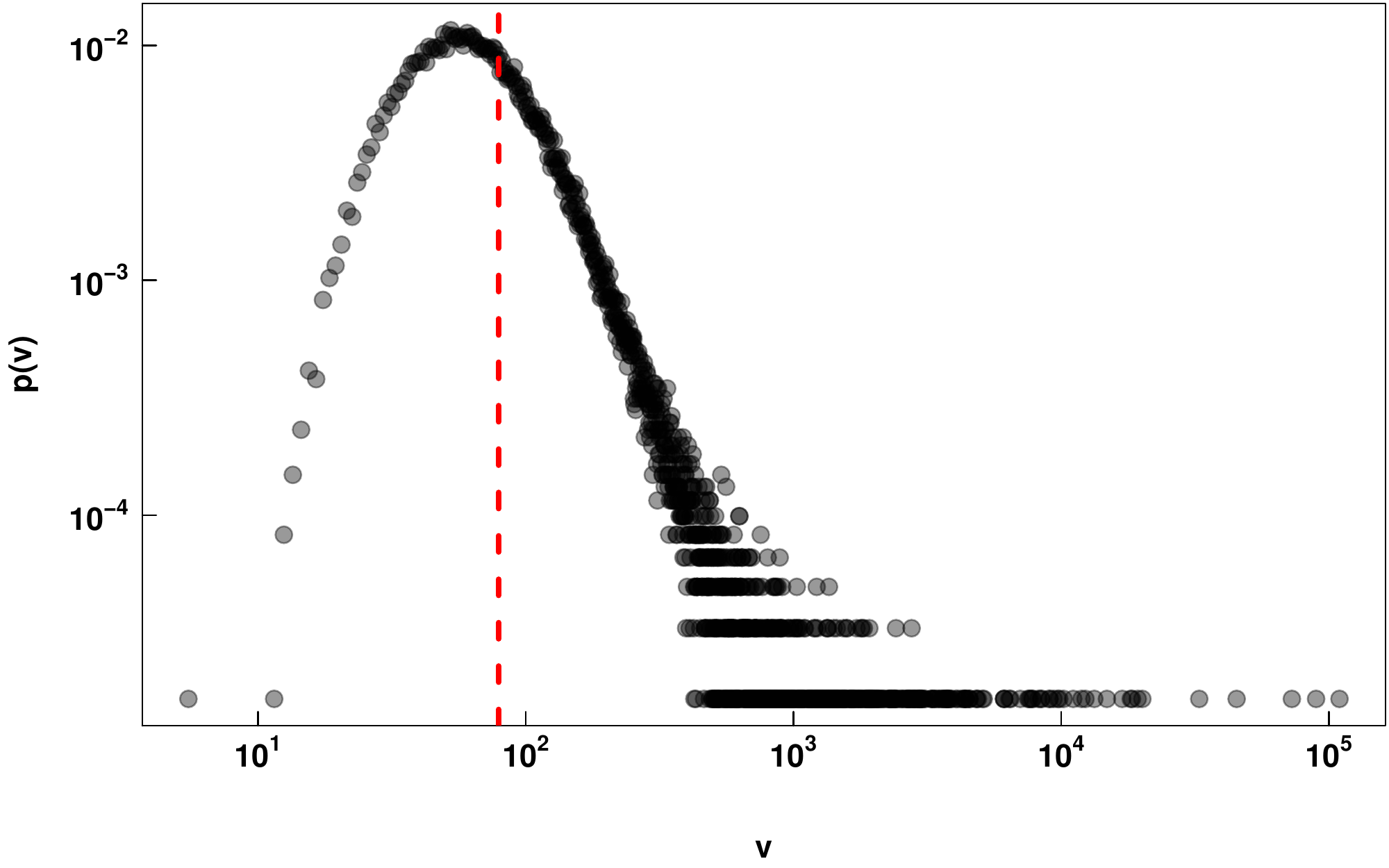}
      \caption{Probability distribution $p(v)$ of the number of views $v$ (log-log scale used). The median value (marked as a dashed line) is around 79 views.}
      \label{fig:views}
\end{figure}

The reason for the above-mentioned fact comes directly from the character of distribution of views depicted in Fig. \ref{fig:views}: it is a heavy-tailed function, however, unlike popular power-law distributions, often encountered in sociophysics, econophysics and science of science \cite{newman2004} it possesses a peak located close to the median value. It immediately gives rise to a crucial question: which papers should be treated as ``popular'' ones taking into account the number of views as a metric? Pursuing this line of thought even further, one might wonder which threshold $v_{th}$ that divides the papers into two classes: popular (say ``positive'') and not popular (say ``negative'') can lead to a good prediction of such classes based on the defined features. 

\section{\textsc{METHODS}}
Data Mining (DM) or Machine Learning (ML) \cite{hastie2001,machinelearning} can be, very generally, described as a rather multidisciplinary field that shows how to learn from data, and make predictions about them. Its connections to physical sciences, which are bi-directional as both disciplines are mutually gaining due to cooperation have lately been emphasised in a recent review \cite{Carlen2019}. The major advantage of DM is that its methods can simultaneously search relations among model variables (features) and the modelled outcome regardless of the size of the space features.    

One of the main categories of ML is \textit{supervised learning} which concerns learning from a set of previously labelled data. In our case, as described in Sec. \ref{sec:sec2}, we would like to to able to predict the class of the paper (popular/not popular), which is a typical example of binary classification. In such a setting, a confusion matrix is a 2x2 table (see Table \ref{tab:tab2}), that allows visualization of the performance of an algorithm, basing on the number of occurrences when originally positive / negative case was truly (or falsely) predicted as positive / negative.

\begin{table}[h!]
\renewcommand*{\arraystretch}{1.4}
\begin{center}
\begin{tabular}{l|c|c|}
\cline{2-3}
 & actual positive & actual negative\\
\hline
predicted positive & true positive (TP) & false positive (FP)\\
\hline
predicted positive & false negative (FN) & true negative (TN)\\
\hline
\end{tabular}
\end{center}
\caption{Confusion matrix}\label{tab:tab2}
\end{table}

However it is usually much better to work on a single index instead of a compound metrics such as a matrix. In this work two such metrics were used: $F_1$ score \cite{f1} i Matthews correlation coefficient (MCC) \cite{mcc} defined as follows
\begin{equation}\label{eq:f1}
F_1 = \frac{2 TP}{2 TP + FP + FN}
\end{equation}
\begin{equation}\label{eq:mcc}
MCC = \frac{TP\cdot TN - FP \cdot FN}{\sqrt{(TP + FP)(TP + FN)(TN + FP)(TN + FN)}}
\end{equation}
with $F_1 \in \left[0, 1\right]$ and $MCC \in \left[-1, 1\right]$. $F_1$ score is a harmonic mean of the precision (i.e., $\frac{TP}{TP+FP}$) and recall (i.e., $\frac{TP}{TP+FN}$). On the other hand MCC is more informative than other confusion matrix measures because it takes into account the balance ratios of the four confusion matrix categories. The value of \textit{MCC} equal to zero means that the model in question chooses class randomly; the closer value to 1 the the better the model performs.

The whole dataset was randomly split into a training and testing part in proportion 75\% to 25\%. Based on Fig. \ref{fig:views} the popularity threshold for ML models was chosen (20 -- 300 views) which covers a sufficiently large range of divisions between popular not not popular papers (see Fig. \ref{fig:percent_of_popular}). For each such value, a set if binary classifiers (see below) was created and their performance was assessed on the testing set.

\begin{figure}[h!]
      \centering
      \includegraphics[width=0.75\linewidth]{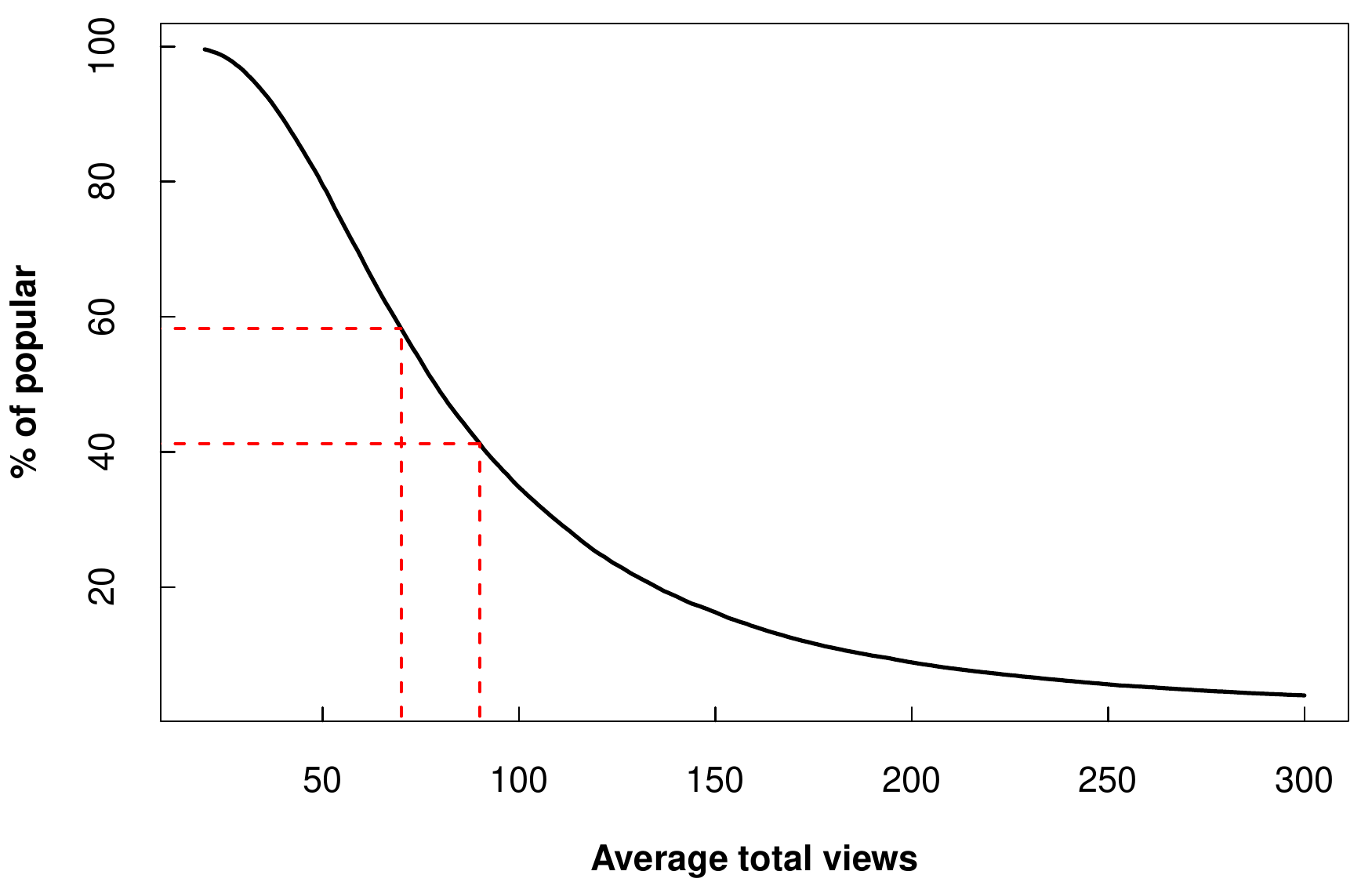}
      \caption{The percentage of popular articles in the data set depending on the popularity threshold. The dashed lines mark the range of 70-90 of views and the corresponding ratios of popular articles.}
      \label{fig:percent_of_popular}
\end{figure}

To compare models with different complexity we chose the following classification algorithms: (i) Linear discriminant analysis (LDA) \cite{fisher_lda, lda_2} projects high-dimensional data onto a low-dimensional space where the data achieves maximum class separability, (ii) QDA, relatively similar to LDA, however, does not assume the covariance of each of the classes to be identical, (iii) Support Vector Machines (SVM) \cite{svm_article} map the data into a higher dimensional input space and construct an optimal separating hyperplane in this space. Due to the kernel method SVM allows for creating nonlinear decision surfaces, (iv) Logistic regression \cite{logistic_regression} is a mathematical modeling approach that can be used to describe the relationship of several features to a dichotomous dependent variable, such as if an article is popular or not, (v) Random forest classifier \cite{random_forest} consists of a combination of tree classifiers where each classifier is generated using a random vector sampled independently from the input vector (features), and each tree casts a vote for the most popular class to classify an input vector. 

\begin{figure}[h!]
\begin{tabular}{cc}
\includegraphics[width=.5\linewidth]{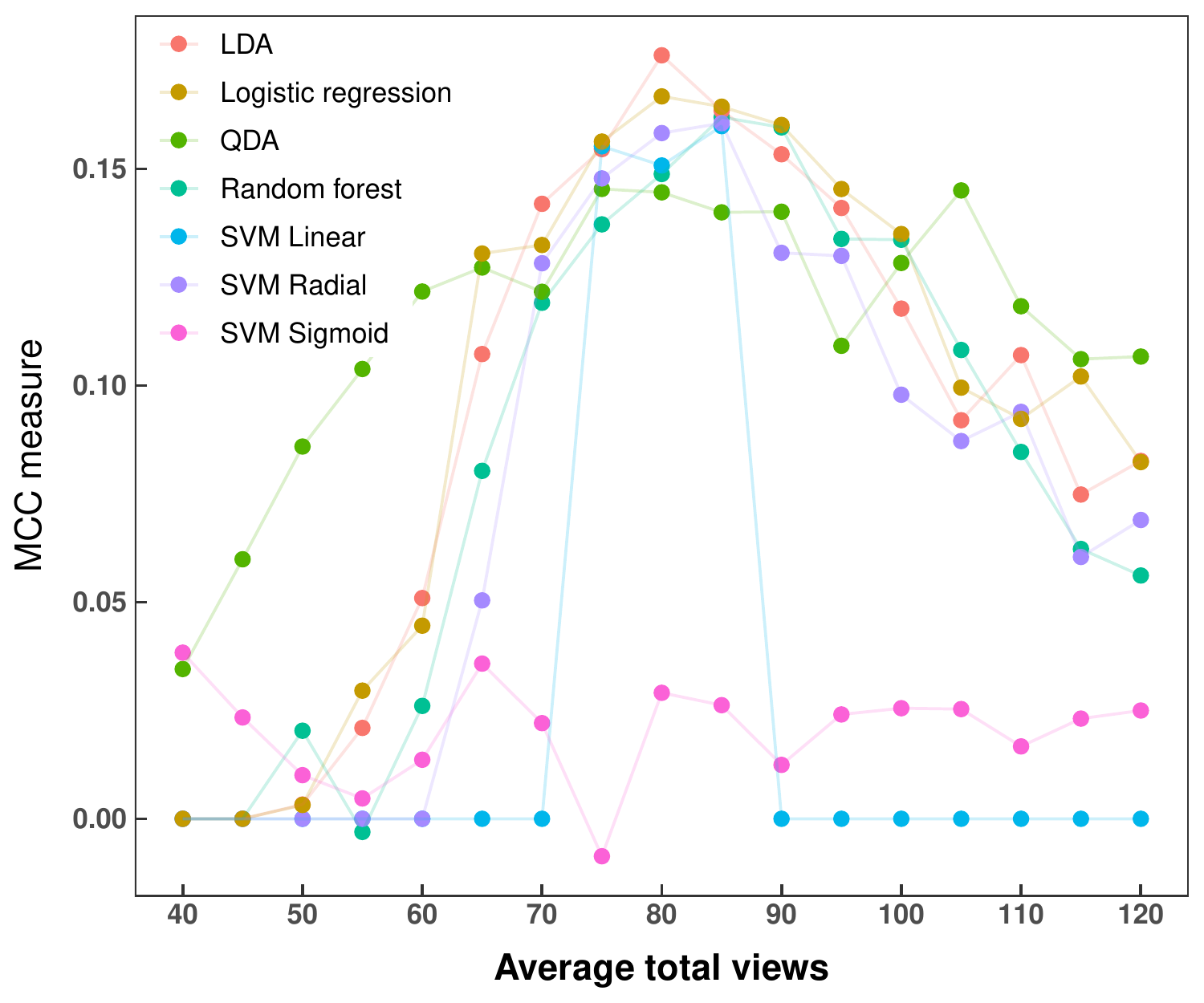} & \includegraphics[width=.5\linewidth]{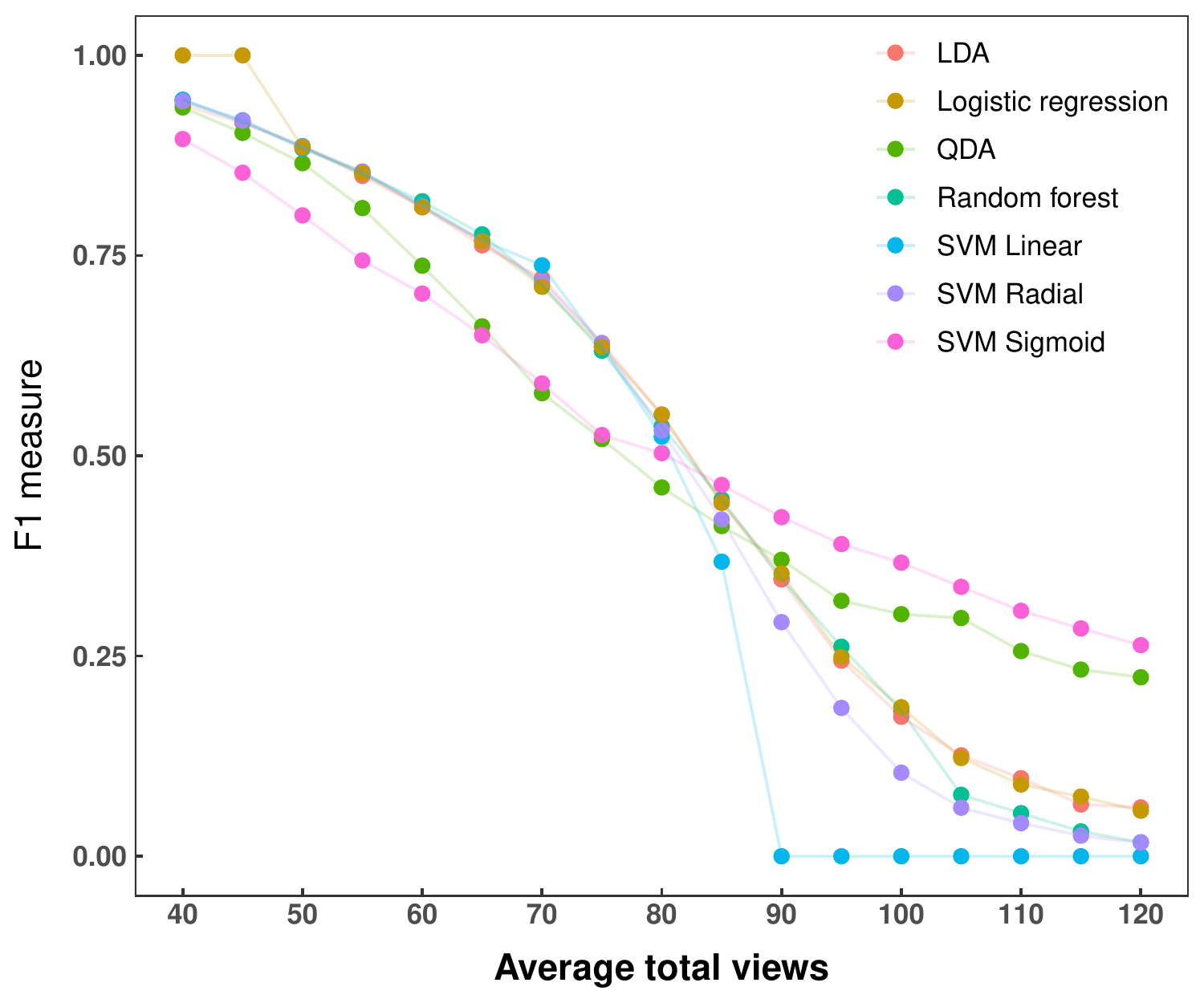}
\end{tabular}
      \caption{(left) \textit{MCC} value versus the threshold of average number of views. (right) $F_1$ score versus the threshold of average number of views.}
      \label{fig:fig4}
\end{figure}

\section{\textsc{RESULTS}}

Figures \ref{fig:fig4} shows the results of each model's \textit{MCC} and $F_1$ values, respectively, as a function of the views threshold for a relevant range of values. It is easy to notice that $F_1$ (Fig. \ref{fig:fig4} - right panel) brings hardly any insights as it is basically monotonically decreasing function with the threshold value. This effect is probably caused by a not balanced dataset -- as it was mentioned before, $F_1$ take into account only $TP$, $FP$ and $FN$ values without the input from $TN$. Thus, for further analysis, we disregard the outcomes of $F_1$. On the other hand, $MCC$ (Fig. \ref{fig:fig4} - left panel) is much more informative: it is clear that the majority of used ML models (except for the sigmoidal version of SVM) tend to have a maximum between 70 and 90 views which roughly corresponds to 40\% -- 60\% popularity (see Fig. \ref{fig:percent_of_popular}). Although the models come from different domains the differences among them are rather minimal with LDA being the best one. One needs to underline that the maximal value of $MCC \approx 0.18$ suggests rather mild predictive power of the approach. Almost all the methods point to $v \approx 80$ as the best threshold value, thus suggesting a 50:50 division between popular and not popular papers. QDA seems to outperform other methods in case of searching for the best model for different regimes of $v$.

\begin{figure}[h!]
      \centering
      \includegraphics[width=.6\linewidth]{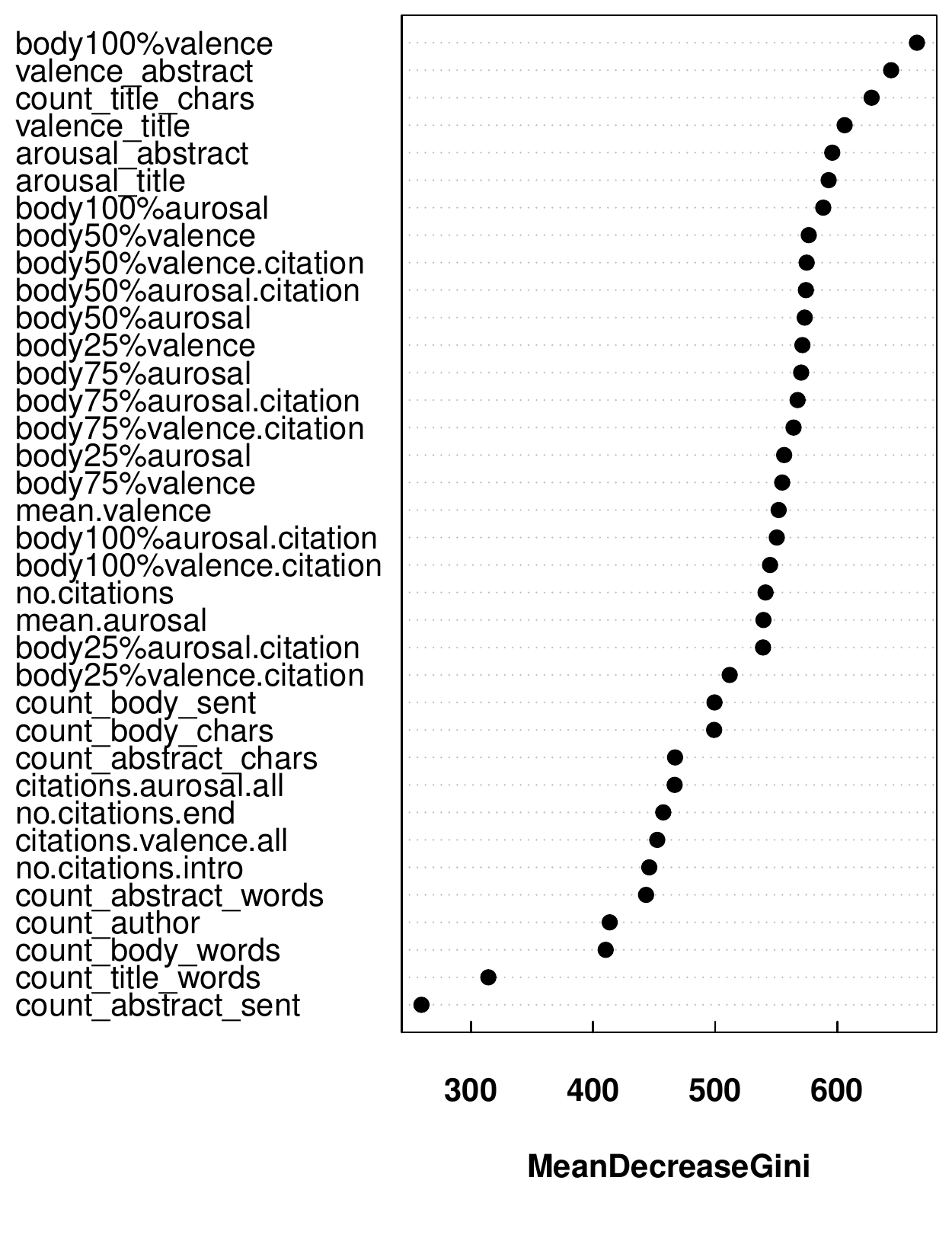}
      \caption{The value of \textit{Mean Decrease Gini} of each feature. The random forest model was built for 80 views.}
      \label{fig:gini}
\end{figure}

Apart from just acquiring the best possible prediction rate, one usually is also interested in understanding primary drivers for such behavior. If the model performs in the same or nearly the same way while some of its features are removed one can conclude that the remaining features are \textit{crucial}, i.e., they are key variables deciding on the modeled outcome. Such an approach is generally referred as to dimension reduction -- its probably most known example is Principal Component Analysis (PCA, described in detail e.g. in \cite{Gajewski2016}) that allows to remove highly correlated variables and with respect to the science of science can be used to show distinctions among scientific areas or fields \cite{Sienkiewicz2018}. However, as one of the methods used in this study is the random forest, we are able to employ \textit{Mean Decrease Gini} instead. Gini Index \cite{breiman_gini} is an attribute selection measure, which measures the impurity of an attribute with respect to the classes. This metric allows to distinguish most influential features for random forest classifier and is often used to reduce the number of features in the input vector (cf. \cite{gini}).



\begin{figure}[h!]
      \centering
      \includegraphics[width=0.98\linewidth]{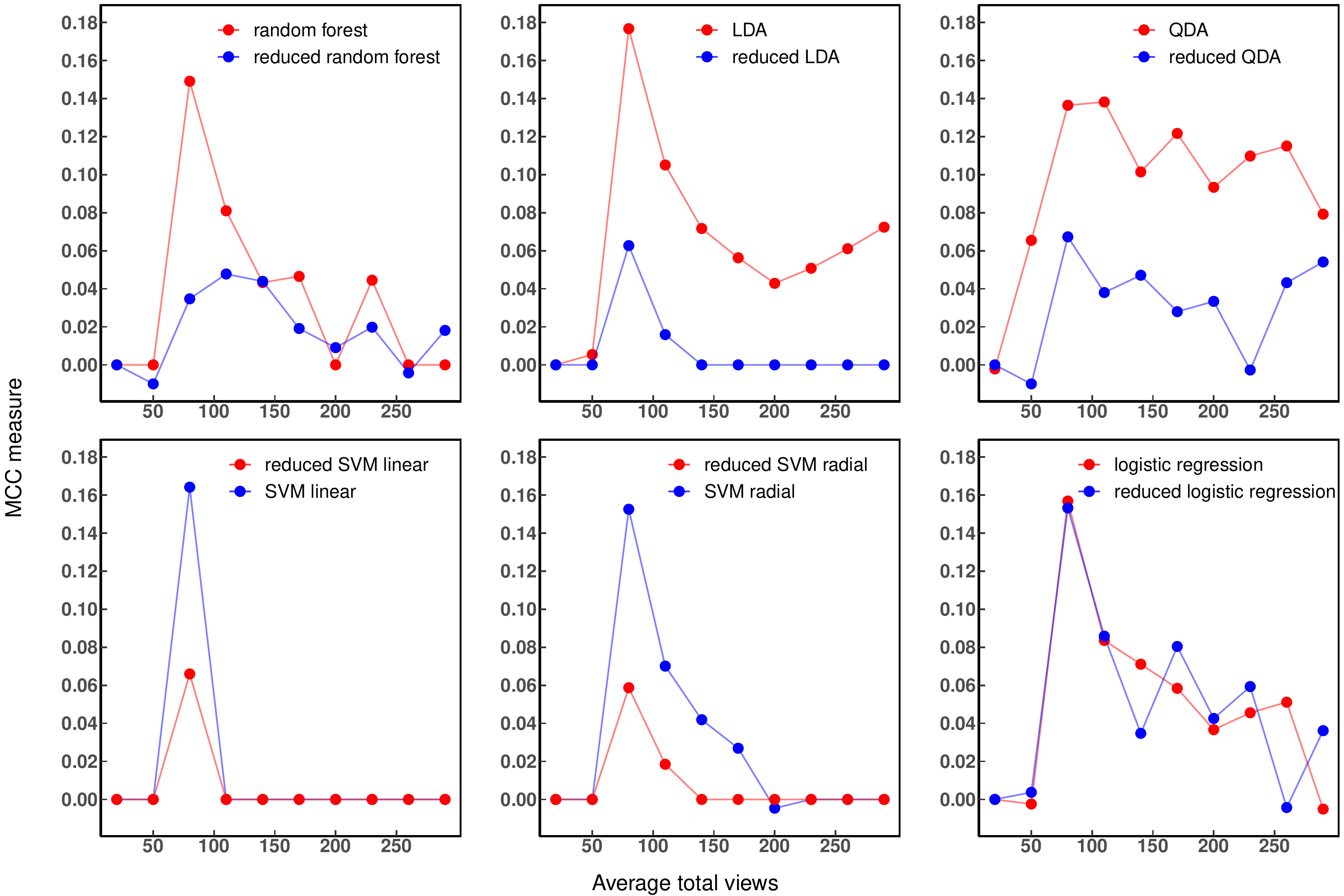}
      \caption{Comparison of \textit{MCC} values for models built on all features and the ones selected by dimension reduction.}
      \label{fig:mcc_zoom}
\end{figure}

We selected the random forest model for the best \textit{MCC} value (80 views). 
Figure \ref{fig:gini} shows the \textit{Mean Decrease Gini} for each feature:
the first three features (i.e., valence in the 4th part of the full text, valence in the abstract as well as the number of characters in the title) were taken to build once again selected machine learning models. A comparison of the models constructed on reduced features and full features is shown in Fig. \ref{fig:mcc_zoom}. For the majority of the models, dimension reduction leads to a drastic decrease of $MCC$ measure, thus implying that the reduced dataset has no predictive power. However, surprisingly in the case of logistic-regression (bottom-right panel in Fig. \ref{fig:mcc_zoom}) both full and reduced sets acquire almost the same $MCC$ value.


\section{\textsc{CONCLUSIONS}}

In this study we have conducted an analysis of over 70.000 articles coming from PLOS One journal, trying to link their lexical, sentiment and bibliographic properties with the popularity they acquired, measured as the number of views of a single paper. In contrast to previous works we have put the emphasis on the question of predicting the popularity and testing the quality of such methods, which has led us to machine learning approaches. In order to overcome the problem of ambiguity of the popularity we have decided to use classification approach by introducing a threshold that divides the number of views and, consequently, the papers themselves into two classes: popular and not popular ones. Basing on the selection of machine learning models the popularity threshold was set between 70 and 90 views with the maximum value close to 80, being the median of the views distribution. For such a threshold the majority of the examined models achieved the best predictive power. Using random forest model we have selected three most crucial features, e.g., valence in the 4th part of full text and in abstract as well as the number of characters in the title. Models built on a reduced number of features performed in general worse than on the total number of features with the prominent exception of logistic regression.

Let us underline that these results do not contradict the conclusions shown in \cite{Sienkiewicz2016} where the strongest identified factors were number of authors and length of abstract (in Fig. \ref{fig:gini} counted as almost least important ones). The difference comes from the fact that in our study we considered a single journal whereas \cite{Sienkiewicz2016} aggregated results from over 1.500 journals (not even including PLOS One). On the other hand our study also brings the length of the title to the front.

In our opinion the study can be easily extended by, e.g., examining the results of Principal Components Analysis or comparing the outcomes of quantile regression method. Moreover as the service of Public Library of Science gives access also to the number of PDF downloads it would be interesting to combine these measures together with the number of citations. 

\bibliographystyle{ieeetr}
\bibliography{jankowski_sienkiewicz_appa_2020}

\end{document}